\documentclass[12pt]{article}
\usepackage{graphicx}

\begin{document}

\title{\hfill {\small CERN-TH/2002-332} \\ \vspace{0cm} \hfill {\small
MCTP-02-60}\\
\vspace{1cm} Observational Tests of Open Strings in Braneworld
Scenarios}

\author{Katherine Freese$^{1,2,3}$, Matthew Lewis$^1$
and Jan Pieter van der Schaar$^{4}$\\
\\
$^1$ Michigan Center for Theoretical Physics,\\  
Physics Department, University of Michigan, \\
Ann Arbor, MI 48109, USA\\
\\
$^2$ Kavli Institute for Theoretical Physics, \\
University of California, Santa Barbara, CA 93106, USA\\
\\
$^3$ ISCAP, Columbia University, 550 W. 120th St., NY, NY, 10027, USA\\
\\
$^4$ CERN Theory Division, CH-1211 Geneva 23, Switzerland}

\maketitle

\begin{abstract}

We consider some consequences of describing the gauge and matter
degrees of freedom in our universe by open strings, as suggested by
the braneworld scenario. We focus on the geometric effects described
by the open string metric and investigate their observational
implications.  The causal structure of spacetime on the brane is
altered; it is described not by the usual metric $g_{\mu\nu}$, but
instead by the open string metric, that incorporates the
electromagnetic background, $G_{\mu\nu} = g_{\mu\nu} - (2\pi
\alpha^\prime)^2 (F^2)_{\mu\nu}$.  The speed of light is now slower
when propagating along directions transverse to electromagnetic fields
or an NS-NS two form, so that Lorentz invariance is explicitly
broken. A generalized equivalence principle guarantees that the
propagation of {\it all} particles, not just photons, (with the
exception of gravitons) is slower in these transverse directions.  We
describe experiments designed to detect the predicted variations in
the causal structure: interferometric laboratory based experiments,
experiments exploiting astrophysical electromagnetic fields, and
experiments that rely on modification to special relativity. We show
that current technology cannot probe beyond open string lengths of
$10^{-13}$ cm, corresponding to MeV string scales.  Should the
experiments someday be able to observe these effects, one could use
them to determine the string scale.  We also point out that in a
braneworld scenario, constraints on large scale electromagnetic fields
together with a modest phenomenological bound on the NS-NS two-form
naturally lead to a bound on the scale of canonical noncommutativity
that is two orders of magnitude below the string length. By invoking
theoretical constraints on the NS-NS two-form this bound can be
improved to give an extremely strong bound on the noncommutative scale
well below the Planck length, $\sqrt{|\theta|_{max}} < 10^{-35} \,
{\rm cm} \times \left({{\rm TeV} \over {\rm string} \,\, {\rm scale}}
\right)^2 $.

\end{abstract}

\newpage

\section{Introduction}

Although string theory is a very promising theoretical framework for
describing quantum gravity and the unification of particles and their
interactions, it has so far been unsuccessful in making contact with
our observable universe. The purpose of this paper is to look for
observational signatures of string physics. In particular we have in
mind the perspective of the recently proposed braneworld scenario
\cite{braneworld}.  In this scenario, our observable universe is a
three-dimensional surface (3-brane) embedded in a higher dimensional
curved spacetime.  The brane degrees of freedom are described by open
strings that end on the brane. Here the gauge and matter degrees of
freedom of the standard model, all described by open strings
\cite{SM-braneworld}, live on the brane.  Gravitons, which are closed
string modes, propagate in the bulk.  The smallness of gravitational
interactions in our brane universe can be explained either by the size
of compact extra dimensions or by the warping of the space transverse
to the brane (effectively reducing the strength of gravitational
interactions to the Planck scale on our observable brane
\cite{ran-sun}). In the latter setup, it is possible to start out with
a fundamental scale in higher dimensions, in our case the string
scale, equal to a few TeV.  In this paper we consider string scales in
the range TeV to $m_{\rm{pl}} = 10^{19}$ GeV.

The fact that all matter and radiation in a braneworld universe are
described by open string states leads to some interesting
consequences. As pointed out long ago, the effective action for the
massless modes of open strings in a slowly varying $U(1)$ background
is given by the Born-Infeld (BI) action\footnote{The generalisation of
this action for $U(N)$ gauge fields is not known in closed form. In
string theory one is trying to make progress order by order, for the
latest results see \cite{NA-BI} and references therein.}.

{\it Causal Structure of Spacetime:} The most important effect of open
string states for the purposes of this paper is that the causal
structure of our brane is altered; it is described not by the usual
metric, $g_{\mu\nu}$, but instead by the open string metric,
$G_{\mu\nu}$ \cite{sei-wit} that incorporates the electromagnetic
background, $F$.  Hence the (3+1)-dimensional 
metric describing propagation in our universe (3-brane) is given by
\begin{equation}
\label{osmetric}
G_{\mu\nu} = g_{\mu\nu} - (2\pi \alpha^\prime)^2 (F^2)_{\mu\nu} \, ,
\end{equation}
where Greek indices $\mu=0,1,2,3$ describe our (3+1)-dimensional spacetime.
The anti-symmetric tensor 
$F_{\mu\nu} = {\cal F}_{\mu\nu} - {\cal B}_{\mu\nu}^{NS}$ is constructed out 
of the electromagnetic field tensor on the brane ${\cal F}=dA$ and the NS-NS 
anti-symmetric 2-form gauge potential ${\cal B}_{\mu\nu}^{NS}$ induced from 
the bulk. The parameter $\alpha^\prime$ represents the squared 
string length which we take to be in the range
\begin{equation}
10^{-33}\, {\rm cm} < l_s \equiv \sqrt{\alpha^\prime} < 10^{-17}\,
{\rm cm} \, . \label{stringlength} 
\end{equation} 
Here the lower limit corresponds to a string scale at $m_{\rm pl}$
while the upper limit corresponds to a TeV string scale.

As we will show, the speed of light is now slower when propagating
along directions transverse to electromagnetic fields.  Assuming that
the ${\cal B}_{\mu\nu}^{NS}$ contribution\footnote{Not to be confused with 
the magnetic field vector $\vec{B}$ or magnitude $B^2$ in the rest of the 
paper.} vanishes, we find the following
expression for the speed of light along directions at an angle
$\theta$ from the magnetic and/or electric field vectors ($\vec{E} ||
\vec{B}$)
\begin{equation}
\label{eq:newc}
{\bar{c} \over  c_{\rm vac}} = \sqrt{  {1-(2\pi \alpha^\prime)^2 E^2 \over 
1 + (2\pi \alpha^\prime)^2 (B^2 \sin^2{\theta} - E^2 \cos^2{\theta})}} \,
\, .\label{cz1}
\end{equation}
For the specific case of $\theta=\pi/2$,
the modified speed of light is
\begin{equation}
\label{apx}
\bar{c} = c_{\rm{vac}} \sqrt{ \frac{1-(2\pi\alpha^\prime)^2E^2}{1
+ (2\pi\alpha^\prime)^2B^2}} \approx [1-2(\alpha^\prime \pi)^2
(E^2+B^2)]c_{\rm{vac}}
\end{equation}
\noindent where the $E$ and $B$ fields are at right angles to the
propagating light, and the approximation is made in the limit of small
$\alpha^\prime \, E$ and $\alpha^\prime \, B$. 

Note that the component of light propagation parallel to the
electromagnetic field is unaffected; hence, we are explicitly breaking
Lorentz invariance in an interesting way.  As light emerges from a
point source in the presence of strong electromagnetic fields, the
surface defined by the distance it has traveled in a given time (the
horizon) is no longer a spherical shell, but rather has the shape of
an ellipsoid (even after the light leaves the vicinity of the fields,
it retains this ellipsoidal shape).

We emphasize that the propagation of {\it all} particles (with the
exception of gravitons) is slower in directions perpendicular to
electromagnetic fields \cite{gib-her}: It is the causal structure of
spacetime on the brane that is modified, and hence the behavior of all
matter and radiation is altered. This effect is a property of
spacetime and not of photons.  One way to visualize this effect is
shown in Figure 1: the lightcone is ``pinched'' in certain directions;
{\it i.e.}, the conventional 45 degree angle between the spatial and
temporal axes becomes smaller.  This pinching takes place in all
directions other than parallel to the electromagnetic field.


\begin{figure}[htb]
\begin{center}
\includegraphics[scale=.40]{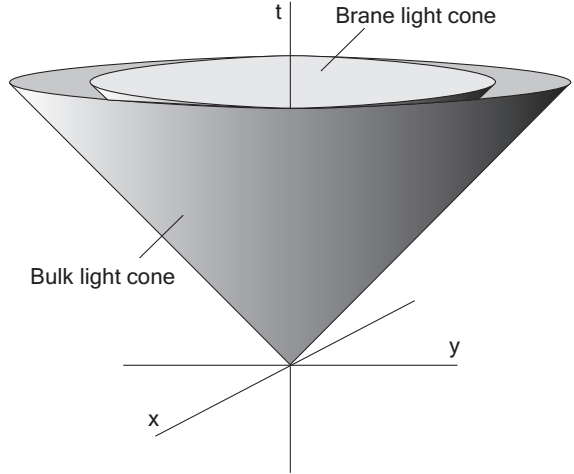}
\end{center}
\caption{The pinching of light cones in the presence of
electromagnetic fields.  The pinching occurs in all directions not
parallel to the field.  Here the fields are confined to the (x-y)
plane and point out of the page.}
\label{fig1}
\end{figure}

We describe several experiments designed to search for this effect. In
general, this effect is suppressed because it is proportional to the
string length to the fourth power, as can be seen in Eq.(\ref{apx}).
We will show that current experiments are able to achieve sensitivity
to a string length of $10^{-13}$ cm, corresponding to a string scale
of 100 MeV.  Hence, many orders of magnitude of improvement in
technology or entirely new methods would be required to observe
anything interesting.  We will discuss (i) interferometric laboratory
based experiments, (ii) experiments exploiting astrophysical
electromagnetic fields, and (iii) experiments that rely on
modification to special relativity.  Because strong EM fields modify
the causal structure of spacetime, classical tests of relativity are
altered; in particular, we examine the fractional change in the muon
lifetime when immersed in a strong magnetic field.

We also note that, although nonlinear quantum electrodynamics (QED)
produces similar but much larger effects, they are in principal
distinguishable from the geometric effects described here.  First,
nonlinear QED applies only to photons, whereas the effects described
in this paper are a result of changes in spacetime itself: these
spacetime effects apply equally to all particles (other than
gravitons) and produce corrections, {\it e.g.} to special relativity,
that cannot possibly be produced by QED.  In addition, because these
spacetime effects do not give rise to birefringence, whereas the
nonlinear QED effects do, they could in principle be differentiated
even in the case of photons.

{\it Gravitons:} The only particles not slowed in the presence of
electromagnetic fields are gravitons.  Gravitons live in the bulk
rather than on the brane and do not feel the previously described
changes in causal structure.  Hence their propagation time, {\it
e.g.}, from a supernova explosion with strong electromagnetic fields,
will be faster than the propagation time of all other particles.  As
discussed below, one can hope to detect the time delay between the
arrival times of neutrinos and gravitons from supernovae.  Such delay
effects were also considered by Chung and Kolb \cite{chungkolb} in the
context of Lorentz violations in braneworlds and by Csaki, Erlich, and
Grojean \cite{csaki-grojean} in the similar context of asymmetrically
warped spacetimes. Following a somewhat different approach, Burgess
{\it et al.} \cite{burgess} have explored constraints on the graviton
dispersion relations arising from Lorentz violations in the bulk and
how such violations are ultimately transmitted to fermion and photon
dispersion relations.

Chung and Freese \cite{cf2} studied a different effect causing the
effective speed of light to be modified.  They proposed ``shortcut
metrics'', whereby geodesics traversing the extra dimensions can allow
communication between points on the brane that are naively causally
disconnected.  They \cite{cf2} proposed these metrics as an
alternative to inflationary scenarios as a solution to the horizon
problem.  Subsequent work on shortcut metrics includes
\cite{caldlang}, \cite{ishihara}, \cite{davis}, and \cite{cf3}.
Shortcut metrics and electromagnetic fields share the feature of
having propagation of signals via the bulk faster than signals
remaining on the brane.  However, shortcut metrics have the end effect
of allowing light signals to appear to travel faster than usual (by
traversing the bulk) while electromagnetic fields have the effect of
slowing down light signals.

{\it Noncommutative Geometry:}

As emphasized by Seiberg and Witten \cite{sei-wit} in the context of
open string theory, an electromagnetic background can be related to
the noncommutativity of spacetime,
\begin{equation}
[x^\mu, x^\nu ] = i \theta^{\mu\nu} \,
\end{equation}
via
\begin{equation}
\theta^{\mu\nu} = 2\pi \alpha^\prime \left( -2\pi \alpha^\prime
F^{\mu \rho} (G^{-1})^\mu_\rho \right) \, , 
\label{noncompar}
\end{equation}
where $\theta^{\mu\nu}$ is the noncommutativity parameter and
$G^{\mu\nu}$ is given in Eq.(\ref{osmetric}).  Astrophysical
electromagnetic fields together theoretical constraints on the NS-NS
2-form ${\cal B}_{\mu\nu}^{NS}$ can then be used to place extremely
strong bounds on the parameter $\theta^{\mu\nu}$.  The absence of
cosmological electric fields of any significant amplitude makes any
numerical bound on the timelike component $\theta^{0i}$ so extremely
tiny that we conclude it is essentially vanishing.  From astrophysical
magnetic fields $B$, we can present an explicit numerical bound on the
scale of spacelike noncommutativity $\theta^{ij}$. For magnetic
backgrounds, we find
\begin{equation}
|\theta| = \sqrt{2} \, (2\pi \alpha^\prime)^2 \, |B| \, . 
\end{equation}
The typical size of intergalactic magnetic fields is roughly
$|B|_{IG} \leq 10^{-9}$ Gauss, so that
we find the following upper bound on the length scale of spacelike 
noncommutativity
\begin{equation}
\sqrt{|\theta|_{max}} \leq 10^{-35} \,  \left[{{\rm string scale}
 \over {\rm TeV}} \right]^{-4} \, {\rm cm} \, . 
\end{equation}
This is an amazingly strong bound on the scale of noncommutativity.
Even without invoking the theoretical constraint on the 2-form ${\cal
B}_{\mu\nu}^{NS}$, but instead using a more modest phenomenological
bound on the scale of ${\cal B}_{\mu\nu}^{NS}$ based on the absence of
strong (anisotropic) effects on the causal structure of spacetime, we
still find that the scale of noncommutativity has to be below the
string length.

{\it Outline:} The organization of this paper is as follows; we begin
by considering open strings ending on D-branes, concentrating on the
$U(1)$ part of the dynamics giving the Born-Infeld action for the
massless modes of the strings.  We next introduce the open string
metric, briefly explain its relevance for finding decoupled
noncommutative field or string theories, and show exactly how the open
string metric affects the causal structure on the brane. In the
following section we describe experiments designed to detect the
nonlinear effects predicted by the open string metric. Finally, we
describe a strong bound on the magnitude of the noncommutativity
parameter by using experimental data on large scale electromagnetic
fields.  We end with some conclusions.

\section{D-branes and the open string metric}
 
Our working assumption will be that our $4$-dimensional universe is a
(wrapped) D-brane embedded in a higher dimensional curved
spacetime\footnote{We will use a mostly plus convention for the
4-dimensional metric, {\it i.e.} we will use
$diag(g_{\mu\nu})=(-1,+1,+1,+1)$ .}, as suggested by the braneworld
scenario. All degrees of freedom confined to the D-brane, and
therefore our universe, are described by open string states. For the
present discussion we can limit ourselves to the subset of bosonic
gauge degrees of freedom ({\it i.e.} photons), because everything we
will conclude about the causal structure can be generalized to include
all open string degrees of freedom.  We refer to \cite{strings-EM} for
a more complete discussion of strings in electromagnetic backgrounds.
When the background fields vary slowly enough\footnote{{\it i.e.} the
variation of the field, $\Delta F$, over a distance $\Delta x$
satisfies $l_s^3 {\Delta F \over \Delta x} \ll 1$.} with respect to
the string length $l_s$, the massless bosonic modes of the string
(photons) can be shown to obey equations of motion that can be deduced
from the following effective Born-Infeld Lagrangian \cite{born-infeld}
\begin{equation}
{\cal L}_{BI} = T \, \sqrt{-{\rm det} \, (g_{\mu\nu} + (2\pi \alpha^\prime) 
F_{\mu\nu}}) \, ,
\label{LBI}
\end{equation}
where $T$ is the brane tension\footnote{For D3-branes the tension can
be written as $T={1 \over (2\pi \alpha^\prime)^2 \, g_s}$, where $g_s$
is the string coupling.} and $g_{\mu\nu}$ is the induced metric on the
brane, defined by the bulk metric $\gamma_{KL}$ and the embedding
scalars $X^K$ as $g_{\mu\nu} = \partial_\mu X^K \partial_\nu X^L
\gamma_{KL}$. Here, Greek indices $\mu = 0, 1, 2, 3$ and Latin indices
$K= 0,1,2,3, ... $ extend also over the extra dimensions.  The
anti-symmetric tensor $F_{\mu\nu} = {\cal F}_{\mu\nu} - 
{\cal B}_{\mu\nu}^{NS}$ is constructed out of the electromagnetic field 
tensor on the brane ${\cal F}=dA$ and the NS-NS anti-symmetric 2-form gauge 
potential ${\cal B}_{\mu\nu}^{NS}$ induced from the bulk. This particular 
combination is the only one invariant under the string worldsheet gauge 
transformations in the presence of D-branes.

It will be important in later sections to distinguish between the pure
electromagnetic contribution that can be probed or detected with
charged open string states ({\it e.g.} electrons), and the NS-NS
2-form contribution to the BI action. Charged open string states
correspond to strings stretching between two different branes $i$ and
$j$ and couple electromagnetically to ${\cal F}^i - {\cal F}^j$.  In
the simplest case of two branes with two $U(1)$'s on the two seperate
brane worldvolumes, the relative difference ${\cal F}^i - {\cal F}^j$
corresponds to the $U(1)$ electrodynamics that we are interested in,
whereas the symmetry under shifts of the overall position of the
D-brane system corresponds to the trivial decoupled $U(1)$. Charged
states cannot be used to measure the ${\cal B}_{\mu\nu}^{NS}$ field,
because they do not couple to it. Even though this is true,
theoretically the ${\cal B}_{\mu\nu}^{NS}$ field is not an independent
or unconstrained field, because the equations of motion following from
(\ref{LBI}) relate it to ${\cal F}$, implying that there can be no
large ${\cal B}_{\mu\nu}^{NS}$ independent of ${\cal F}$. In fact, in
realistic brane world models \cite{SM-braneworld} one typically
introduces orientifold planes where the boundary condition at a plane
sets the NS-NS 2-form ${\cal B}_{\mu\nu}^{NS}$ to zero\footnote{We
would like to thank Joseph Polchinski for very helpful discussions and
correspondence on the distinction and relation between the NS-NS
2-form ${\cal B}_{\mu\nu}^{NS}$ and the electromagnetic ${\cal F}$.}.
This will be important to keep in mind in the following sections when
we discuss the electromagnetic effects on the causal structure or the
noncommutativity parameter, because in principle there could be an
independent contribution from the NS-NS 2-form ${\cal
B}_{\mu\nu}^{NS}$.

To lowest order in $\alpha'$, the nonlinear BI Lagrangian can be
expanded to give the standard Maxwell action. To see this we use the
fact that the 4-dimensional determinant in the BI Lagrangian $-{\rm
det} \, (g_{\mu\nu} + (2\pi \alpha^\prime) F_{\mu\nu}) \equiv (-{\rm
det} \, g_{\mu\nu})\, D$ can be written as follows
\begin{eqnarray}
D &\equiv& {\rm det} \left( \delta^{\rho}_{\nu} + (2\pi \alpha^\prime) 
{F^{\rho}}_{\nu} \right)  \nonumber \\
&=& \left( 1-{1\over 2} 
(2\pi \alpha^\prime)^2 {\rm Tr}\, F^2 + {1\over8} (2\pi \alpha^\prime)^4
({\rm Tr} \, F^2)^2 - {1\over 4} (2\pi \alpha^\prime)^4 {\rm Tr} \,  F^4 
\right) \, , \label{detn+f}
\end{eqnarray}
where we used ${\rm Tr} \, F^2 \equiv g^{\mu\nu}
(F^2)_{\mu\nu}$, ${\rm Tr} \, F^4 \equiv g^{\mu\nu} (F^4)_{\mu\nu}$ and 
(anti-symmetric) matrix multiplication in powers of $F_{\mu\nu}$, so 
$(F^2)_{\mu\nu} \equiv F_{\mu\rho} g^{\rho \delta} F_{\delta \nu}$. 
Expanding the square root in (\ref{LBI}), using (\ref{detn+f}), we find
\begin{equation}
{\cal L}_{BI} = T \, \sqrt{-{\rm det} \, g_{\mu\nu}} \left( 1-{1\over
4} (2\pi \alpha^\prime)^2 {\rm Tr}\, F^2 + {\cal O}({\alpha^\prime}^4)
\right) = T \, \sqrt{-{\rm det} \, g_{\mu\nu}} + {\cal L}_{YM} + T \,
{\cal O}({\alpha^\prime}^4) \, . \label{YMlag}
\end{equation} 

It has long been recognized that the nonlinear BI theory has some
remarkable properties \cite{Schrodinger, Boillat}. In particular, the
propagation of fluctuations around an electromagnetic background
solution has causal properties very different from its linear and
classical Maxwell cousin. From an open string perspective these
massless fluctuations are identified with massless open string states
propagating in a nontrivial background. The study of open string
states propagating in a constant electromagnetic background has
received a lot of attention due to its relation to noncommutative
geometry that was unraveled by Seiberg and Witten \cite{sei-wit}. In
the presence of a nontrivial background the natural open string
parameters are the open string metric, the open string coupling and an
anti-symmetric tensor that can be interpreted as describing the
noncommutativity of spacetime $[x^\mu, x^\nu]=i \, \theta^{\mu\nu}$.
The open string metric and the noncommutativity tensor are
respectively determined by the symmetric and anti-symmetric part of
the propagator relevant for open string vertex operators. Without
going into detail, we present expressions for these open string
parameters \cite{sei-wit}
\begin{eqnarray}
G_{\mu\nu} &=& g_{\mu\nu} - (2\pi \alpha^\prime)^2 (F^2)_{\mu\nu} \, , \\
\lambda_{os} &=& g_s \, \sqrt{{\rm det} \, g_{\mu\nu} + (2\pi \alpha^\prime)
F_{\mu\nu} \over {\rm det} \, g_{\mu\nu}} \, , \label{oscoupling} \\
\theta^{\mu\nu} &=& 2\pi \alpha^\prime \left( -2\pi \alpha^\prime
F^{\mu \rho} (G^{-1})^\mu_\rho \right) \, . 
\end{eqnarray}
When taking the zero slope (or point particle) limit $\alpha^\prime
\rightarrow 0$ of the open string theory in the presence of a
background electric or magnetic field, the crucial observation is that
one should concentrate on the scaling of the open string parameters
(\ref{osmetric}), (\ref{oscoupling}) and (\ref{noncompar}) to see
whether one obtains a nontrivial decoupled theory on the brane. Zero
slope limits were constructed that gave rise to either noncommutative
gauge theories (for magnetic backgrounds) \cite{sei-wit} or
noncommutative open string theories (for electric backgrounds)
\cite{stro-mal, sei-sus} on the brane, decoupled from the bulk
gravitational theory. Throughout the rest of this paper we will be
interested in the spacetime causal structure described by the open
string metric (\ref{osmetric}) and the noncommutativity of spacetime
as described by (\ref{noncompar}), rather than the open string
coupling constant (\ref{oscoupling}).

The appropriate metric for all open string degrees of freedom on the
D-brane, therefore, is $G_{\mu\nu}$, rather than the induced metric
$g_{\mu\nu}$.  The induced metric is the appropriate metric to
describe the bulk fields (particles that are not confined to the
D-brane, {\it e.g.}  gravitons) This fact can also be deduced by
looking at the equations of motion following from (\ref{LBI}) instead
of considering the full open string theory. The Born-Infeld equations
of motion can all be rewritten using the open string metric instead of
the induced metric.  Hence, by turning on electric or magnetic fields
on the brane, we can change the causal structure, or equivalently, we
can change the speed of light ($c_{vac}$). We would however like to
emphasize that the relevance of the open string metric is not limited
to the BI action, but instead describes the causal properties of the
full open string theory.

From an experimental point of view, one might worry because nonlinear
QED produces competing effects that are typically much larger than
those produced in nonlinear BI theory.  However, these two effects can
be distinguished due to the fact that nonlinear BI effects have no
polarization dependence and hence do not exhibit the birefringence (or
bi-metricity, {\it e.g.} \cite{visser}) associated with nonlinear QED.
Second, when we consider the effects due to the full open string
metric, we obtain changes not only to the Lagrangian of
electrodynamics but also changes in the causal structure of spacetime.
The equivalence principle is at work in the case of the open string
metric; all open string states, not just photons, are affected in the
same manner.  Hence, whereas nonlinear QED applies only to photon
interactions, the geometric effects discussed below apply to all
matter and radiation propagating on the brane.

Let us take a more precise look at the causal structure described by
the open string metric \cite{gib-her}. The open string lightcone is
defined as the set of 4-vectors $V^\mu$ satisfying $G_{\mu\nu}V^\mu
V^\nu \equiv 0$.  On the other hand the usual lightcone is defined by
the (null) vectors $W^\mu$ and the induced metric $g_{\mu\nu}$ as
$g_{\mu\nu} W^\mu W^\nu=0$; from now on we will use the terminology
``bulk'' lightcone for the latter.  The quantity
\begin{equation}
G_{\mu\nu} W^\mu W^\nu = - (2\pi\alpha^\prime)^2 (F^2)_{\mu\nu} W^\mu
W^\nu \geq 0 \label{gcone}
\end{equation}
is always positive because the term $(F^2)_{\mu\nu} W^\mu W^\nu \leq
0$ for generic vectors $W^\mu$.  Eq.(\ref{gcone}) implies that the
vector $W^\mu$ is generically spacelike with respect to the open
string metric $G_{\mu\nu}$.  One can also see that the open string
lightcone touches the bulk lightcone along the two principal null
directions of the electromagnetic field tensor $F_{\mu\nu}$, which are
defined by
\begin{equation}
{F_\mu}^\nu W^{\mu} = \lambda W^{\nu} \Rightarrow (F^2)_{\mu\nu} W^\mu
W^\nu =0 \, . \label{eigenvec}
\end{equation}
In other words, in the two directions parallel to the electromagnetic
field, the field $F_{\mu\nu}$ has no effect, so that there is no
difference between $g_{\mu\nu}$ and $G_{\mu\nu}$.  The open string
lightcone touches the bulk lightcone along the two principal null
directions but otherwise lies within the bulk lightcone.

Figure 1 is a plot of both lightcones, with respect to the metric
$g_{\mu\nu}$.  The open lightcone is ``pinched;'' {\it i.e.}, it lies
inside the bulk lightcone, everywhere except along the two principal
null directions, where the two lightcones touch.  This pinching of the
lightcone is a different result than one would get from a change in
spacetime curvature; curved spacetime has light cones that are tilted
rather than pinched.  The open string metric therefore does not curve
the spacetime with respect to the bulk metric; rather, it affects the
local speed of light in all directions other than the special
principal null directions.  At the risk of repeating ourselves, it is
important to emphasize that this result is not just restricted to the
speed of photons, but pertains to all propagating open string states
and it is in that sense that a generalized equivalence principle is at
work here. The degree pinching of the lightcone depends on the size of
the electromagnetic background, the string length, and the direction
under consideration.

Let us now concentrate on the electromagnetic contribution to this
effect and explicitly identify the electric and magnetic field
backgrounds in the standard way ($i,j \in (1,2,3)$)
\begin{equation}
E_i \equiv F_{0i} \quad , \quad B^i \equiv {1\over 2} 
\epsilon^{ikl} F_{kl} \, , \label{EBident}
\end{equation}
where $\epsilon^{ikl}$ is the (Euclidean) 3-dimensional anti-symmetric
Levi-Cevita tensor with $\epsilon^{123}=1$. Assuming a flat induced
metric, {\it i.e.} $g_{\mu\nu}=\eta_{\mu\nu}$, we obtain the following
proper distance from the open string metric (\ref{osmetric})
\begin{eqnarray} ds^2 = G_{\mu\nu} dx^\mu dx^\nu = &-& \left[ 1-(2\pi
\alpha^\prime)^2 E^2 \right] dt^2 \nonumber \\ &-& (2\pi
\alpha^\prime)^2 (\vec{E} \cdot d\vec{x})^2 + (2\pi \alpha^\prime)^2
(\vec{E} \times \vec{B}) \cdot d\vec{x} \, dt \\ &+& \left[ 1 + (2\pi
\alpha^\prime)^2 B^2 \right] d\vec{x}^2 - (2\pi \alpha^\prime)^2
(\vec{B} \cdot d\vec{x})^2 \nonumber \, .
\label{properdis}
\end{eqnarray}
For illustrational purposes let us simplify the situation further and
assume that the $\vec{E}$ and $\vec{B}$ vectors are parallel (so
$\vec{E} \times \vec{B}$ vanishes) and introduce $\theta$ as the angle
between the propagation direction $d\vec{x}$ and $\vec{E} \|
\vec{B}$. We then find
\begin{eqnarray}
ds^2 = &-& \left[ 1-(2\pi \alpha^\prime)^2 E^2 \right] dt^2
\nonumber \\ &+& \left[ 1 + (2\pi \alpha^\prime)^2 B^2
\sin^2{\theta} - (2\pi \alpha^\prime)^2 E^2 \cos^2{\theta} \right]
d\vec{x}^2 \, ,
\label{coordxyz}
\end{eqnarray}
which clearly shows that the speed of light $c_{vac}$ is only affected
when propagating along directions in the plane orthogonal to the
magnetic and electric field ($\theta= \pi /2$), as explained before in
more general terms. Along the magnetic and electric field, the
principal null direction of $F_{\mu\nu}$ ($\theta=0$), the speed of
light is not affected.  We are thus breaking Lorentz invariance in an
unusual manner by turning on a magnetic or electric field. To
summarize, we find the following expression for the speed of light
along directions at an angle $\theta$ from the magnetic and/or
electric field vectors ($\vec{E} \| \vec{B}$)
\begin{equation}
{\bar{c} \over  c_{\rm vac}} = \sqrt{  {1-(2\pi \alpha^\prime)^2 E^2 \over 
1 + (2\pi \alpha^\prime)^2 (B^2 \sin^2{\theta} - E^2 \cos^2{\theta})}} \,
\, .
\label{cz}
\end{equation}
As concluded before, this shows that the open string lightcone lies
within the bulk, or closed string, lightcone.  Thus the speed of light
can only become smaller. If it would have been the other way around
the speed of light would be unbounded from above (depending on the
direction and the electromagnetic field), leading to problems with
causality. It is also worth pointing out that in open string theory
and BI theory there exists a maximal, critical, electric field that is
obtained as $E$ approaches ${1\over 2\pi \alpha^\prime}$. At this
critical electric field, the speed of light vanishes.

From the open string perspective there exists an intuitive way to
understand these effects. The open string endpoints carry Chan-Paton
factors, which are essentially charges with respect to the gauge
fields on the brane. In the $U(1)$ case with a non-trivial background
the open string is like a dipole rod and therefore wants to line up
with the field, which affects its orientation and its effective
tension effectively changing the causal structure in the way we just
described.

In the following sections we will estimate the generic magnitude of
this effect and describe possible experiments to detect it. We will
try to make full use of the predicted equivalence principle and
emphasize the open string nature of this effect.

\section{Detecting changes in causal structure}

In this section, we explore experimental avenues for observing the
changes in causal structure predicted by Eq. \ref{cz}. We first
consider methods for inducing variations in the speed of light due to
electromagnetic backgrounds, and investigate the feasibility of
detecting such changes using available technology. If these effects
were observed, one could effectively measure the string length. From
Eq. \ref{cz}, the modified speed of light, $\bar{c}$, in the presence
of electric and magnetic fields transverse to the propagation
direction,

\begin{equation}
\bar{c} = c_{\rm{vac}} \sqrt{ \frac{1-(2\pi\alpha^\prime)^2E^2}{1 +
(2\pi\alpha^\prime)^2B^2}} \approx [1-2(\alpha^\prime \pi)^2
(E^2+B^2)]c_{\rm{vac}}
\end{equation}

\noindent where the fields $E||B$ are at right angles to the
propagating light ($\theta=\pi/2$), and the approximation is made in
the limit of small $\alpha^\prime$: an optimistic, TeV scale string
length would imply $\sqrt{\alpha^\prime} = 1.98 \times 10^{-17}$ cm.

In the absence of any electromagnetic contributions, the effect we are
looking for could be due to a NS-NS two-form ${\cal B}_{\mu\nu}^{NS}$
that except for noncommutative signatures that we will discuss later,
would be undetectable. The absence of a large anisotropy in the causal
structure, which would result in variations in the (local) speed of
light in different directions, can therefore be used to constrain the
NS-NS two-form ${\cal B}_{\mu\nu}^{NS}$.

\subsection{Laboratory experiments: Interferometers}

Do we have any hope of seeing a change in the speed of light in a
laboratory experiment? Detecting small variations in the speed of
light is probably best approached with interferometric methods. To
this end, we examine an experiment involving an idealized
interferometer. A laser is split into two beams; one beam travels in a
vacuum, and the other travels in a region immersed in strong,
transverse electric and magnetic fields.  Because the light beam
moving in the electromagnetic fields propagates slower, the two beams
acquire a relative phase shift.  When the light beams are recombined,
their phase difference may be measured using interferometric methods;
roughly, the intensity of the recombined light beams is a measure of
the relative phase difference.  By modulating the electromagnetic
fields in the interferometer, we induce a time dependent phase shift
that would be essential to achieving a viable signal to noise ratio.

To estimate the sensitivity required of such an interferometer,
consider two light waves, $\psi_1$ and $\psi_2$, with the former
propagating {\it in vacuo}, and the latter traveling in transverse
electromagnetic fields,

\begin{eqnarray}
\psi_1 = A\sin(2\pi f t - k_1 x)\\
\psi_2 = A\sin(2\pi f t - k_2 x)
\nonumber
\end{eqnarray}

\noindent where $f$ is the common frequency of the waves, $k_1 = 2\pi
f/c_{\rm{vac}}$ and $k_2 = 2\pi f/\bar{c}$.  After each of the light
waves has traveled a distance $L$, the relative phase difference
accumulated is,

\begin{eqnarray}
\Delta \phi = k_2 L - k_1 L = \frac{2\pi L }{\lambda_{\rm{vac}}}
\left(\sqrt{ \frac{1+(2\pi\alpha^\prime)^2B^2}{1 -
(2\pi\alpha^\prime)^2E^2}} - 1 \right) \approx \frac{4\pi
(\pi\alpha^{\prime})^2 L}{\lambda_{\rm{vac}}} (E^2 + B^2)
\label{phaseshift}
\end{eqnarray}

\noindent where $\lambda_{\rm{vac}}$ is the vacuum wavelength of the
light, and the ratio $\bar{c}/c_{\rm{vac}}$ has been determined by
Eq.(\ref{cz}).  To assess the feasibility of this technique, we
estimate the expected phase shift under favorable experimental
circumstances.

To achieve the largest change in phase, we want to employ the largest
available electric and magnetic fields.  As large electric fields are
easier to modulate than large magnetic fields, they are preferred.
High electric fields on the scale of $10^{9}$ V/m can be produced, but
such fields pose a number of technical challenges.  The ionization of
residual gas in the vacuum and the electric ``puncture'' of the
dielectric material encasing the vacuum limit the practical size of
the field.  In the context of this experiment, electric field
strengths on the order of $10^8$ V/m should be achievable.  The
largest DC magnetic fields available for controlled terrestrial
experiments are on the order of $45$ Tesla \cite{magnet}.  Pulsed
electromagnets can provide fields somewhat higher ({\it e.g.}  $70$
Tesla) but would be unsuitable for use in sensitive interferometric
experiments.

The relative phase shift is also proportional to the total path length
of the two beams.  The further the beams propagate, the more time they
have to accumulate phase difference. In principle, the interferometer
could be several kilometers in length ({\it e.g.} LIGO, \cite{ligo}),
but sustaining high electromagnetic fields on such scales would be
challenging, to say the least.  Alternatively, an array of coupled,
high-finesse Fabry-Perot interferometers arranged in a compact
geometry could allow for an effective path length as high as $10^7$
cm. In this scenario, a laser is split, and the resulting beams are
sent into separate Fabry-Perot cavities, one of which is isolated from
EM fields, and one of which is exposed to high EM fields. In the
Fabry-Perot cavities, the beams undergo multiple reflections ($\sim
10^6$, for extremely high finesse cavities).  This process can be
repeated many times, effectively yielding a large path length for the
beams. Because this path length is compressed into a small region, the
problem of producing large EM fields is made relatively easier.

With these considerations, and assuming a TeV scale string length,
$\sqrt{\alpha^\prime} = 1.98 \times 10^{-17}$ cm, the most auspicious
phase shift is,

\begin{eqnarray}
\label{estimate}
\Delta \phi = 2 \times 10^{-26}\mbox{ rad} \times \left(
\frac{\alpha^\prime}{3.9 \times 10^{-34} \mbox{ \rm{cm}}^2}\right)^2
\left( \frac{L}{1.0 \times 10^6 \mbox{ cm}}
\right)\left(\frac{\lambda_{\rm{vac}}}{10^{-4} \mbox{ cm}}
\right)^{-1}\\ \nonumber \times \left[ \left( \frac{E}{1.4 \times 10^{10}
\mbox{ V/m}}\right)^2 + \left( \frac{B}{45 \mbox{ T}}\right)^2 \right]
\end{eqnarray}

Optical phase detection is currently achievable at the $10^{-10}$ rad
level \cite{lantz}; phase sensing in this regime is quantum limited by
the statistics of photon detection and limitations in beam intensity.
Although detection sensitivity can be improved by increasing the
intensity of the laser, distortion of the optics due to thermal
heating places a practical limit on beam power.  Methods are being
developed to cope with these problems, but even if future techniques
improve this limit by a few orders of magnitude, we still fall at
least 12 orders of magnitude short of detecting this effect. A
photon-based experiment using ring lasers has been proposed by Denisov
in the context of classical BI theory, but is comparable to the above
interferometric experiment in its sensitivity \cite{denisov}.  Even
more problematic is the fact that when we restrict ourselves to
photons the effect will be entirely swamped by QED interactions that
produce a similar effect (but with bi-refringence) at the much lower
scale of the electron mass \cite{nlQED}. In fact, a proposal to
measure the nonlinear QED effect in the near future has appeared
recently \cite{boerholten}. This unwanted competition can be avoided
by considering neutral (nearly) massless particles different from
photons. Of course, such particles are typically more difficult to
produce and control than photons.

Given current limitations in achievable EM field strength and detector
sensitivity, interferometric and ring laser experiments could detect
nonlinear electromagnetic effects only if the string length was of
order $\sqrt{\alpha^\prime} \sim 10^{-13}$ cm, corresponding to an MeV
string scale.  Hence, experiments using earth-bound EM fields to
detect variations in the speed of light for string scales above TeV
currently seem unfeasible.

\subsection{Astrophysical bounds}

To help save the situation, we can look for larger electromagnetic
fields, possibly in astrophysical and cosmological contexts.
Extremely intense magnetic fields are found near highly magnetized
pulsars, known as magnetars; magnetars can exhibit fields exceeding
$10^9$ Tesla.  Although this would increase the observed phase shift
by some 16 orders of magnitude over that predicted in
Eq.(\ref{estimate}), it is not clear how one could exploit these large
(but remote) fields to make sensitive phase measurements.  Similarly
large electric fields exist near such objects, but present the same
experimental challenges.

Alternatively, one could try to make use of the fact that
gravitational fluctuations are unaffected by the open string metric;
instead, their propagation is governed by the usual closed string
metric because it is a bulk closed string excitation.  Hence, for an
astrophysical event (such as a supernova or a neutron star inspiral)
that releases copious amounts of gravitational radiation and
neutrinos, there may be a measurable time delay between the observed
arrival of the gravitational radiation and the neutrinos.  Situations
where gravitons travel with a speed different than the speed of light
on the brane have also been studied in the context of asymmetrically
warped spacetimes \cite{csaki-grojean}.  In the present case, the
delay could be a result of the intense electromagnetic fields
associated with the astrophysical event itself, or a result of the
intervening galactic magnetic fields between the event and the
earth. To see if this effect would be measurable, we note that for an
event occurring a distance $L$ from the earth, the expected time delay
is on the order of,

\begin{equation}
\Delta t = L \left[\frac{1}{\bar c} - \frac{1}{c}\right] \approx
\frac{2(\pi\alpha^\prime)^2 B^2 L}{c}
\end{equation}

Galactic magnetic fields, coherent on $1-2$ kpc scales, exist but are
typically very weak, $\sim10^{-10}$ Tesla.  For an event a distance $L
= 10$ Mpc from the earth, and assuming the intervening galactic
magnetic fields were optimally oriented, the expected time difference
is

\begin{equation}
\Delta t \approx 10^{-43} \, \rm{s}
\end{equation} 

\noindent If one considers the 100 Gauss magnetic field that may
surround supernovae to distances of $10^{16}$ cm, the delay is
somewhat improved.  In such a situation, the delay is now,

\begin{equation}
\Delta t \approx 10^{-24} \, \rm{s}
\end{equation}

\noindent In either case, the time difference falls ridiculously short
of detectability.  With foreseeable LIGO technology, a delay of at
least several seconds, and more probably several days would be
required to reliably observe such an effect. There may be other
processes better suited to inducing a delay, but present astrophysical
conditions do not seem to provide a satisfactory environment for
studying variable speed of light effects.

\subsection{Special relativistic effects: Lifetime of the muon}

Experiments that do no rely explicitly on measuring variations in the
speed of light are also possible.  Changing the speed of light would
impact the classical tests of special relativity by changing the usual
relativistic $\gamma$ factor.  For example, the lifetime of a muon
moving in a strong EM field would be slightly longer than that of a
muon traveling at the same speed in a vacuum.  The expected fractional
change in the muon lifetime can be estimated,

\begin{equation}
\label{frchng}
\frac{\Delta \tau}{\tau} = \frac{\tau_0 \bar{\gamma} -
\tau_0\gamma}{\tau_0
\gamma}=\left({\frac{1-\frac{v^2}{c_{\rm{vac}}^2}}{1-\frac{v^2}
{\bar{c}^2}}}\right)^{1/2}-1 \approx
\frac{2v^2}{c_{\rm{vac}}^2(1-v^2/c_{\rm{vac}}^2)} (\pi\alpha^\prime)^2
[E^2 + B^2]
\end{equation}

\noindent where $\tau_0$ is the lifetime of the muon in its rest
frame, $v$ is the speed of the muon, and where Eq.(\ref{apx}) was used
in the last step.  For a 2 GeV muon, with the same optimistic
estimates of string length and EM field strength made in
Eq.(\ref{estimate}), Eq.(\ref{frchng}) predicts a fractional change in
lifetime of,

\begin{equation}
\frac{\Delta \tau}{\tau} = 1.4 \times 10^{-34}
\end{equation}

Precision experiments can measure the muon lifetime only to $1$ part
in $10^6$; hence, tests exploiting relativistic phenomena in this
manner also do not appear to be practical.

\subsection{Constraining ${\cal B}_{\mu\nu}^{NS}$ contributions}

We have focused on electromagnetic contributions that change the
causal structure of spacetime. In principle, however, there could also
be an independent contribution from the NS-NS 2-form ${\cal
B}_{\mu\nu}^{NS}$.  A large, constant, NS-NS ${\cal B}_{\mu\nu}^{NS}$
would distort the causal structure of spacetime, inducing an
observable anisotropy in the speed of light. Thus, to the extent that
the speed of light is observed to be isotropic, we can constrain the
magnitude of a possible NS-NS ${\cal B}_{\mu\nu}^{NS}$ contribution.

A uniform ${\cal B}_{\mu\nu}^{NS}$ would single out a preferred
direction in the universe along which light would propagate at
$c_{\rm{vac}}$.  Light propagating perpendicular to this preferred
direction would move more slowly. The difference in the observed
speeds may be estimated using equation \ref{cz}:

\begin{equation}
\Delta c = (c_{\rm{vac}} - \bar c) \approx c_{\rm{vac}} -
c_{\rm{vac}}[1-2(\pi\alpha^\prime)^2 |{\cal B}^{NS}|^2] \, .
\end{equation}

\noindent so that,

\begin{equation}
2(\pi\alpha^\prime)^2 |{\cal B}^{NS}|^2 \le \frac{\Delta
c}{c_{\rm{vac}}}.
\end{equation}

\noindent Modern Michelson-Morley type experiments have constrained
the speed of light to be isotropic to within $\Delta c = 0.9$ m/s
$\sim 3\times 10^{-9}$ c$_{\rm{vac}}$ \cite{riis}.  This in turn
constrains the maximum scale of $|{\cal B}_{\mu\nu}^{NS}|$,

\begin{equation}
(2 \pi \alpha^\prime) \, |{\cal B}^{NS}| \le 8 \times 10^{-5} \, .
\label{Bbound}
\end{equation}

\noindent That is, the NS-NS 2-form must be many orders of magnitude
smaller than the inverse squared string length.

\section{Noncommutativity in a braneworld}

So far we have emphasized the effect that electromagnetic backgrounds
have on the causal structure of spacetime, based on the assumption
that our universe is a collection of branes with the open strings
ending on the branes describing all the Standard Model matter and
gauge degrees of freedom. Under the same assumption it is natural to
relate an electromagnetic background to a noncommutativity parameter,
through the relation (\ref{noncompar}).  There has been a lot of
activity recently in identifying and constraining noncommutative
physics \cite{NC-QL,
NC-LE}\cite{Mazumdar-Jabbari}-\cite{Chaichian:2000si} (for a more
complete list of references we refer the reader to the review
\cite{NC-rev}).  Low energy effects, based on the noncommutative
breaking of Lorentz invariance, provide a very strong bound at the
$10^{14}$ GeV level, or $\sqrt{|\theta|} < 10^{-28}$ cm \cite{NC-LE}.
On a more theoretical note it seems that evaluating loop integrals
without a momentum space cutoff, due to dangerous UV/IR mixing, gives
rise to severe problems in noncommutative theories which can be used
to exclude these noncommutative theories altogether \cite{NC-QL}.
Introducing a momentum space cutoff softens these problems, but strong
bounds can still be constructed that constrain the length scale of
noncommutativity to be a lot smaller than the cutoff length
\cite{NC-QL}.  What all these approaches have in common is that they
treat the noncommutativity parameter as an independent, free variable
and constrain its magnitude through the high and low energy (quantum)
effects of noncommutativity.  As we will discuss, however, from a
stringy braneworld perspective it is more natural to treat the
noncommutativity tensor as a dependent parameter which will allow us
to put a strong (classical) bound on the scale of noncommutativity.

As was explained in \cite{sei-wit}, the effective low energy physics
of open string theory in an electromagnetic\footnote{What we will mean
with an electromagnetic background is something that can be detected
and distinguished from the NS-NS 2-form by the use of charged open
string states. In the absence of any charged particles to measure
electromagnetic fields, a NS-NS 2-form and electromagnetic background
are completely equivalent.} or NS-NS 2-form background can be
described either by commutative, ordinary, gauge fields or by
noncommutative gauge fields, depending on how you regularize the gauge
field interactions on the open string worldsheet. Similar conclusions
hold when one considers charged open string with their ends ending on
different branes \cite{NCOS-charged}. In the commutative description
one keeps the nontrivial Lorentz violating electromagnetic or NS-NS
background that one started with as part of the low energy effective
theory, whereas in the noncommutative description the background is
replaced by a noncommutativity tensor that similarly breaks the
Lorentz symmetry. The precise recipe for replacing an electromagnetic
or NS-NS background by a noncommutativity tensor is given by
Eq.(\ref{noncompar}).

We will assume that this correspondence continues to hold in more
complicated open string brane models that are able to reproduce the
Standard Model at low energy \cite{SM-braneworld}. Under that
assumption we conclude that when a nontrivial electromagnetic or NS-NS
background is present one should be able to reformulate the low energy
physics in terms of a noncommutativity tensor and the corresponding
noncommutative gauge fields, with the relation between the
electromagnetic background and the noncommutativity tensor given by
(\ref{noncompar}). So in this context any such background can be
replaced with a noncommutativity tensor and vice versa; they are not
independent observables. It is important to realize that this
correspondence relies crucially on the existence of an underlying
(open) string theory, {\it i.e.} it involves higher order string
length corrections. One way to see this is by looking at the relation
between electromagnetic or NS-NS backgrounds and the noncommutativity
parameter (\ref{noncompar}) and observing that it degenerates in the
zero-slope limit $\alpha^\prime \rightarrow 0$.

Let us investigate the relation (\ref{noncompar}) more precisely.
We will be interested in the explicit dependence of the
noncommutativity parameter on the field $F_{\mu\nu}={\cal F}_{\mu\nu}-
{\cal B}_{\mu\nu}^{NS}$, which is
not immediately obvious because of the presence of the inverse open
string metric $G^{-1}$, defined through $G_{\mu\rho}
(G^{-1})^{\rho\nu}=\delta_{\mu}^{\nu}$. Using some properties of
4-dimensional anti-symmetric tensors one can show that the inverse
open string metric equals
\begin{equation}
(G^{-1})^{\mu\nu} = g^{\mu\nu} + {D_0 \over D} (F^2)^{\mu\nu} + {1 \over D}
(F^4)^{\mu\nu} \, , \label{inverse-osm}
\end{equation}
where $D$ is given by (\ref{detn+f}) and we have introduced $D_0$ defined as 
\begin{equation}
D_0 \equiv  1-{1\over 2} (2\pi \alpha^\prime)^2 {\rm Tr}\, F^2 \, .
\label{Dnot}
\end{equation}
To continue it will be useful to concentrate on the magnitude of the
noncommutativity parameter and typically one would therefore calculate
the scalar invariant ${\rm Tr}\, \theta^2 \equiv g_{\mu\nu}
(\theta^2)^{\mu\nu}$. However, from the brane perspective it makes
more sense to calculate the scalar invariant with respect to the open
string metric instead of the bulk, closed string metric. Although the
difference for small fields is higher order in the fields and
therefore negligible, for large fields it makes an important
difference.  We will instead define and calculate ${\rm Tr}_G \,
\theta^2 \equiv g_{\mu\nu} (\theta^2_G)^{\mu\nu}$, where the subscript
$G$ implies that traces are taken with respect to the open string
metric on the brane. In fact this calculation is easy because the
inverse open string metric appears in the definition of
$\theta^{\mu\nu}$ (\ref{noncompar}). We find the following result
\begin{equation}
{\rm Tr}_G \, \theta^2 = (2\pi \alpha^\prime)^4 \, {\rm Tr} \, F^2 \, .
\label{theta2}
\end{equation}
Defining the magnitude of the noncommutativity parameter as 
$|\theta| \equiv \sqrt{|{\rm Tr}_G \, \theta^2|}$ we can write 
\begin{equation}
{|\theta| \over (2\pi \alpha^\prime)} = (2\pi \alpha^\prime) 
\sqrt{{\rm Tr} \, F^2} \, ,
\label{thetaM}
\end{equation}
We see that the magnitude of the noncommutativity parameter is not
bounded in principle, but it is clear that extremely large fields
($(2\pi \alpha^\prime) \sqrt{{\rm Tr} \, F^2} > 1$) are required to
get noncommutativity scales larger than the string length. It is
exactly this large field limit that Seiberg and Witten discuss in
their paper \cite{sei-wit} to obtain decoupled noncommutative field
theories. In this limit it is crucial to consider traces with respect
to the open string metric, because the effects on the open string
causal structure (\ref{osmetric}) will now be very
important\footnote{It can be shown that from the perspective of the
bulk metric the size of the noncommutativity parameter never exceeds
the string length.}.

We will use the relation (\ref{thetaM}) to determine a strong bound on
the scale of noncommutativity. In principle, there are two independent
contributions to the noncommutativity parameter. However, as alluded
to earlier, there are theoretical reasons to think that either the
NS-NS 2-form contribution is zero because of the presence of
orientifold planes, or the NS-NS 2-form contribution is similar in
magnitude to the electromagnetic component via the BI equations of
motion. In that case, one is forced to argue that the electromagnetic
contribution is either larger than or comparable to the ${\cal
B}_{\mu\nu}^{NS}$ component. We can then use bounds on the size of
large scale, cosmological, electromagnetic fields to construct a very
strong bound on the noncommutativity parameter. Indeed, cosmological
electric fields are essentially absent to incredible precision and
this will make any numerical bound on the timelike noncommutative
component so extremely tiny that we conclude it has to be
vanishing. Because magnetic field magnitudes are relatively well known
on large scales, we can present an explicit numerical bound on the
scale of spacelike noncommutativity. We find from (\ref{thetaM})
\begin{equation}
{|\theta| \over (2\pi \alpha^\prime)} = \sqrt{2} \, (2\pi \alpha^\prime) 
\, |B| \, . \label{thetaB}
\end{equation}
To obtain a conservative bound on an average noncommutative parameter,
we use the typical size of intergalactic magnetic fields, which is
roughly $|B|_{IG} \leq 10^{-9}$ Gauss; in natural units this number
corresponds to $10^{-2}$ cm$^{-2}$. The average magnetic field in the
universe is probably much smaller than this number so that this
estimate is conservative.  As before, we take the TeV length scale of
about $10^{-17}$ cm as the largest possible string scale.  Then we
find the following upper bound on the length scale of spacelike
noncommutativity
\begin{equation}
\sqrt{|\theta|_{max}} < 10^{-35} \, {\rm cm} \times \left({{\rm TeV} \over
{\rm string} \,\, {\rm scale}} \right)^2 .
\end{equation}
Assuming a minimum string scale of around a TeV, this bound can also
be written as
\begin{equation}
{\sqrt{|\theta|_{max}} \over l_s} \leq 10^{-18} \, . \label{ubound}
\end{equation}
This is an rather straightforward and very strong bound on the scale
of noncommutativity.  Even though there is some room to play with the
numbers, it is clear that the average scale of noncommutativity in a
braneworld universe can only be extremely small, if one assumes the
theoretical constraints on the NS-NS 2-form ${\cal B}_{\mu\nu}^{NS}$
as discussed previously.  This bound then implies that the length
scale of noncommutativity will always be many orders of magnitude
below the Planck length.  Turning things around, any sizeable scale of
noncommutativity bigger than the string length would lead to huge
electromagnetic backgrounds that would have been detected already.

If one does not assume any theoretical constraints on the NS-NS
2-form, one could imagine, as is typically done, that there exists a
large uniform ${\cal B}_{\mu\nu}^{NS}$ background, which would only be
detectable through its noncommutative {\it and} its pinched lightcone
effects on the braneworld geometry. To obtain a noncommutative
parameter comparable to the string scale it is clear from
(\ref{thetaM}) that one needs very large field contributions that
would leave a clear imprint on the causal structure of spacetime that
would most likely, due to its anisotropic nature, have been detected
already. As explained in section 3.4 one can therefore put a
phenomenological bound on the NS-NS 2-form based on local causal
structure constraints (\ref{Bbound}). Because of (\ref{thetaM}) even
this rather modest bound on the NS-NS 2-form will ensure that the
noncommutativity parameter does not exceed the string length. To be
precise we obtain from (\ref{Bbound}) and (\ref{thetaM}) the purely
phenomenological bound
\begin{equation}
{\sqrt{|\theta|_{\rm phen}} \over l_s} \leq 10^{-2} \, . \label{uphenobound}
\end{equation}
Hence we conclude that in the context of braneworld scenarios, one
should concentrate on trying to find stringy rather than
noncommutative experimental signatures.

Relating the noncommutativity parameter to background fields makes it
clear that the noncommutative parameter can and typically will be
spacetime dependent. Even if one assumes that the ${\cal
B}_{\mu\nu}^{NS}$ contribution should be considered as a uniform
vacuum expectation value (VEV), the electromagnetic background
certainly varies in spacetime. Indeed, it seems natural to expect that
the NS-NS 2-form will generate a mass term, ensuring that the vacuum
expectation value of the NS-NS gauge potential vanishes \cite{NC-LE},
or is related to the electromagnetic background by the equations of
motion, and any nonzero noncommutativity parameter is predominantly a
result of nonvanishing electromagnetic backgrounds. The
noncommutativity parameter is thus typically a local, spacetime
varying object and not a universal, fundamental parameter. Many of the
bounds from the previous literature implicitly relate the
noncommutativity parameter to a uniform NS-NS two-form VEV and
therefore allow for the interpretation as a bound on a fundamental
quantity.  It seems more natural to assume that the noncommutativity
parameter is defined only locally and can vary in space and time;
thus, experimental bounds should be interpreted with this in mind.
For example, if the bound relies on measurements on the earth, then it
is possible that the fields, either ${\cal F}_{\mu\nu}$ or ${\cal
B}_{\mu\nu}^{NS}$ or both, are much larger elsewhere in the universe.
Hence, the bounds do not constrain any fundamental quantity, but
rather only its value locally wherever the experiment was made.  Such
measurements then become similar to constraining the magnetic field in
a particular room, which doesn't teach us much about the universe.
From that perspective, when we bound the noncommutativity parameter by
using measurements of average, large scale magnetic fields we
constrained only the average large scale size of this parameter in a
braneworld universe; locally our bounds do not have to be
satisfied. Note, however, that even the largest electromagnetic fields
in our universe will not lead to noncommutativity parameters larger
than the string length. On the other hand, when we invoked the
phenomenological bound on the two-form ${\cal B}_{\mu\nu}^{NS}$ and
interpreted it as a uniform VEV, we must expect this bound to apply
even locally instead of on average. Combining these results it seems
safe to conclude that the {\it local} noncommutativity parameter is
already constrained to be smaller than the string length, and on
theoretical grounds is expected to be many orders of magnitude below
that in a realistic braneworld scenario.

\section{Conclusions}

In an attempt to find observational evidence for the existence of open
strings we have concentrated on the distinct geometric effects of open
strings in braneworld scenarios. We conclude that a controlled
experiment capable of measuring the string length is not within reach
any time soon.  We believe, however, that the universal nature of the
geometric effects that we have described may still allow for a clever,
as yet unknown, experimental procedure with adequate accuracy.
Considerations of open strings in background fields have significant
overlap with investigations into Lorentz violating operators,
nonlinear electrodynamics and noncommutative physics, and in some
sense unifies these investigations.

By considering the open string causal structure we were able to
present a purely phenomenological bound on the size of the NS-NS
two-form ${\cal B}_{\mu\nu}^{NS}$. Although the bound is not very
strong it can be used to show that the noncommutativity parameter
cannot exceed the string length. We think this is a simple, classical,
but nevertheless compelling result in line with other investigations
into the scale of noncommutative physics. Invoking theoretical
motivations to relate the two-form ${\cal B}_{\mu\nu}^{NS}$ to the
electromagnetic background on the brane, or setting it to zero by
assuming either the presence of orientifolds or the appearance of a
mass term, noncommutativity in a realistic braneworld can effectively
be ruled out; {\it i.e.}, it must be orders of magnitude below the
string length.

Of course, the conclusions of this work depend upon the assumption
that we live in a particular braneworld scenario, or more precisely
that all matter and gauge degrees of freedom in our universe are
described by open strings. This might not be the case. One could
speculate that similar geometric effects could occur for closed
strings in nontrivial curved gravitational, or other closed string,
backgrounds due to classical string length corrections. The importance
of trying to find accessible experimental signatures that would reveal
the stringy nature of the constituents of our universe, open or
closed, can hardly be overstated.

\section{Acknowledgements}
We would like to thank Joe Polchinski for interesting and very helpful
discussions, in particular on the difference between NS-NS gauge
potentials and electromagnetic backgrounds on the brane. KF would like
to thank Erick Weinberg for pointing out the potential importance of
nonlinear QED as a competing effect. JPvdS would like to thank Adi
Armoni and Zheng Yin for interesting discussions on noncommutativity
and for pointing out useful references. He would also like to thank
the CERN Theory Division for hospitality and support during the final
stages of this work. ML thanks Keith Riles for useful conversations
regarding the limits of interferometric technology.  KF and ML thank
the Michigan Center for Theoretical Physics for support. KF, ML and
JPvdS thank the DOE via the Physics Dept. at the University of
Michigan for support. KF and ML thank the Kavli Institute for
Theoretical Physics for hospitality and support via a grant from NSF.
KF thanks ISCAP (the Institute for Strings, Cosmology, and
Astroparticle Physics) at Columbia University, where part of this work
was done, for hospitality and support during her stay.


\begin{thebibliography}{99}

\bibitem{braneworld}
N.~Arkani-Hamed, S.~Dimopoulos and G.~R.~Dvali,
``The hierarchy problem and new dimensions at a millimeter,''
{\it Phys.\ Lett.\ B} {\bf 429}, 263 (1998)
[arXiv:hep-ph/9803315].
\\
I.~Antoniadis, N.~Arkani-Hamed, S.~Dimopoulos and G.~R.~Dvali,
``New dimensions at a millimeter to a Fermi and superstrings at a TeV,''
{\it Phys.\ Lett.\ B} {\bf 436}, 257 (1998)
[arXiv:hep-ph/9804398].
\\
N.~Arkani-Hamed, S.~Dimopoulos and G.~R.~Dvali,
``Phenomenology, astrophysics and cosmology of theories with  sub-millimeter dimensions and TeV scale quantum gravity,''
{\it Phys.\ Rev.\ D} {\bf 59}, 086004 (1999)
[arXiv:hep-ph/9807344].
\\
Z.~Kakushadze and S.~H.~Tye,
``Brane world,''
{\it Nucl.\ Phys.\ B} {\bf 548}, 180 (1999)
[arXiv:hep-th/9809147].

\bibitem{SM-braneworld}
M.~Cvetic, G.~Shiu and A.~M.~Uranga,
``Three-family supersymmetric standard like models from intersecting  brane worlds,''
{\it Phys.\ Rev.\ Lett.}  {\bf 87}, 201801 (2001)
[arXiv:hep-th/0107143].
\\
R.~Blumenhagen, B.~Kors, D.~Lust and T.~Ott,
``The standard model from stable intersecting brane world orbifolds,''
{\it Nucl.\ Phys.\ B} {\bf 616}, 3 (2001)
[arXiv:hep-th/0107138].
\\
L.~E.~Ibanez, F.~Marchesano and R.~Rabadan,
``Getting just the standard model at intersecting branes,''
{\it JHEP} {\bf 0111}, 002 (2001)
[arXiv:hep-th/0105155].

\bibitem{ran-sun}
L.~Randall and R.~Sundrum,
``An alternative to compactification,''
{\it Phys.\ Rev.\ Lett.}  {\bf 83}, 4690 (1999)
[arXiv:hep-th/9906064].

\bibitem{born-infeld}
M.~Born and L.~Infeld,
``Foundations Of The New Field Theory,''
{\it Proc.\ Roy.\ Soc.\ Lond.\ A} {\bf 144}, 425 (1934).

\bibitem{Schrodinger}
E.~Schrodinger, 
``Contributions to Born's New Theory of the Electromagnetic Field'', 
{\it Proc. Roy. Soc.}  {\bf 150A} (1935) 465. 
\\
E.~Schrodinger, ``Non-Linear Optics'', 
{\it Proc. Roy. Irish. Acad.} {\bf A 47} (1942) 77.
\\
E.~Schrodinger, 
 ``A new exact solution in non-linear optics (two-wave-system)'', 
{\it Proc. Roy. Irish. Acad.} {\bf A 49} (1943) 59.

\bibitem{Boillat} 

G. Boillat, ``Vari\'et\'es caract\'eristiques ou surfaces d'ondes en
\'electrodynamiques non lin\'eaire'', {\it C. R. Acad. Sci. Paris} {\bf 262}
(1966) 1285.
\\
G.~Boillat, ``Surfaces d'ondes compar\'es de la theorie
d'Einstein-Schroedinger et de l'\'electrodynamique non lin\'eaire;
champs absolus'', {\it C. R. Acad. Sci. Paris} {\bf 264} (1967) 113.
\\
G.~Boillat, 
``Nonlinear Electrodynamics: Lagrangians and Equations of Motion'', 
{\it J. Math. Phys.} {\bf 11} (1970) 941.    
\\
G.~Boillat, ``Exact Plane-Wave Solution of Born-Infeld Electrodynamics'', 
{\it Lett. al Nuovo Cimento} {\bf 4} (1972) 274.
\\
G.~Boillat, 
``Shock relations in Non-Linear Electrodynamics'', 
{\it Phys. Lett.} {\bf 40A} (1972) 9.
\\
G.~Boillat, 
``Convexit\'e et hyperbolicit\'e en \'electrodynamique non-lin\'eaire'', 
{\it C. R. Acad. Sci. Paris} {\bf 290} (1980) 259.

\bibitem{NA-BI}
P.~Koerber and A.~Sevrin,
``The non-abelian D-brane effective action through order alpha'**4,''
{\it JHEP} {\bf 0210}, 046 (2002)
[arXiv:hep-th/0208044].

\bibitem{sei-wit}
N.~Seiberg and E.~Witten,
``String theory and noncommutative geometry,''
{\it JHEP} {\bf 9909}, 032 (1999)
[arXiv:hep-th/9908142].

\bibitem{chungkolb} 
D. Chung, E.W. Kolb, and A. Riotto, ``Extra Dimensions Present a New
Flatness Problem'', {\it Phys. \ Rev. \ D} {\bf 65}, 083516 (2002).

\bibitem{cf2} 
D. Chung and K. Freese, ``Can Geodesics in Extra
Dimensions Solve the Cosmological Horizon Problem?'', {\it Phys.\ Rev.} D
{\bf 62}, 063513 (2000).

\bibitem{caldlang} 
R. Caldwell and D. Langlois, ``Shortcuts in the fifth dimension'',
{\it Phys. Lett.} B511 129-135 (2001)

\bibitem{ishihara} H. Ishihara,``Causality of the Brane Universe'',
{\it Phys.\ Rev.\ Lett.} {\bf 86}, 381 (2001).

\bibitem{davis} A.C. Davis, C. Rhodes, and I. Vernon, ``Branes on the
  Horizon'', {\it JHEP} {\bf 0111}, 015 (2001).

\bibitem{cf3} D. Chung and K. Freese, ``Lensed Density Perturbations
in Braneworlds,'' arXiv:astro-ph/0202066 (2002).

\bibitem{strings-EM}
J.~Ambjorn, Y.~M.~Makeenko, G.~W.~Semenoff and R.~J.~Szabo,
``String theory in electromagnetic fields,''
arXiv:hep-th/0012092.

\bibitem{gib-her}
G.~W.~Gibbons and C.~A.~Herdeiro,
``Born-Infeld theory and stringy causality,''
{\it Phys.\ Rev.} D {\bf 63}, 064006 (2001)
[arXiv:hep-th/0008052].

\bibitem{stro-gop}
R.~Gopakumar, S.~Minwalla, N.~Seiberg and A.~Strominger,
``OM theory in diverse dimensions,''
{\it JHEP} {\bf 0008}, 008 (2000)
[arXiv:hep-th/0006062].

\bibitem{stro-mal}
R.~Gopakumar, J.~M.~Maldacena, S.~Minwalla and A.~Strominger,
``S-duality and noncommutative gauge theory,''
JHEP {\bf 0006}, 036 (2000)
[arXiv:hep-th/0005048].

\bibitem{sei-sus}
N.~Seiberg, L.~Susskind and N.~Toumbas,
``Strings in background electric field, space/time noncommutativity  and a new noncritical string theory,''
{\it JHEP} {\bf 0006}, 021 (2000)
[arXiv:hep-th/0005040].

\bibitem{ber-vds}
E.~Bergshoeff, D.~S.~Berman, J.~P.~van der Schaar and P.~Sundell,
``Critical fields on the M5-brane and noncommutative open strings,''
{\it Phys.\ Lett.\ B} {\bf 492}, 193 (2000)
[arXiv:hep-th/0006112].

\bibitem{visser}
M.~Visser, C.~Barcelo and S.~Liberati,
``Bi-refringence versus bi-metricity,''
arXiv:gr-qc/0204017.

\bibitem{hei-eul}
W.~Heisenberg and H.~Euler,
``Consequences Of Dirac's Theory Of Positrons,''
{\it Z.\ Phys.}  {\bf 98}, 714 (1936).

\bibitem{schwinger}
J.~S.~Schwinger,
``On Gauge Invariance And Vacuum Polarization,''
{\it Phys.\ Rev.} {\bf 82}, 664 (1951).

\bibitem{boerholten}
D.~Boer and J.~W.~Van Holten,
``Exploring the QED vacuum with laser interferometers,''
arXiv:hep-ph/0204207.

\bibitem{nlQED}
V.~A.~De Lorenci, R.~Klippert, M.~Novello and J.~M.~Salim,
``Light propagation in non-linear electrodynamics,''
{\it Phys.~Lett.~B} {\bf 482} (2000) 134
[arXiv:gr-qc/0005049].
\\
M.~Novello, V.~A.~De Lorenci, J.~M.~Salim and R.~Klippert,
``Geometrical aspects of light propagation in nonlinear electrodynamics,''
{\it Phys.~Rev.~D} {\bf 61} (2000) 045001
[arXiv:gr-qc/9911085].
\\
M.~Novello and J.~M.~Salim,
``Effective Electromagnetic Geometry,''
{\it Phys.~Rev.~D} {\bf 63} (2001) 083511.
\\
W.~Dittrich and H.~Gies, ``Light propagation in nontrivial QED
vacua,'' {\it Phys.~Rev.~D} {\bf 58}, 025004 (1998) [hep-ph/9804375].
\\
W.~Dittrich and H.~Gies, {\em Probing the Quantum Vacuum\/}, Springer
Tracts in Modern Physics {\bf 166} (Springer, 2000)

\bibitem{NC-rev}
I.~Hinchliffe and N.~Kersting,
``Review of the phenomenology of noncommutative geometry,''
arXiv:hep-ph/0205040.

\bibitem{Mazumdar-Jabbari}
A.~Mazumdar and M.~M.~Sheikh-Jabbari,
``Noncommutativity in space and primordial magnetic field,''
{\it Phys.\ Rev.\ Lett.}  {\bf 87}, 011301 (2001)
[arXiv:hep-ph/0012363].
\\
\bibitem{Kamoshita}
J.~i.~Kamoshita,
``Probing noncommutative space-time in the laboratory frame,''
arXiv:hep-ph/0206223.
\\
\bibitem{Carroll-etal}
S.~M.~Carroll, J.~A.~Harvey, V.~A.~Kostelecky, C.~D.~Lane and T.~Okamoto,
``Noncommutative field theory and Lorentz violation,''
{\it Phys.\ Rev.\ Lett.}  {\bf 87}, 141601 (2001)
[arXiv:hep-th/0105082].
\\
H.~Falomir, J.~Gamboa, M.~Loewe, F.~Mendez and J.~C.~Rojas,
``Testing spatial noncommutativity via the Aharonov-Bohm effect,''
Phys.\ Rev.\ D {\bf 66}, 045018 (2002)
[arXiv:hep-th/0203260].

\bibitem{csaki-grojean} 
C. Csaki, J. Erlich, and C. Grojean, ``Gravitational Lorentz
Violations and Adjustment of the Cosmological Constant in
Asymmetrically Warped Spacetimes'', {\bf Nucl.Phys. B}604 312-342
(2001)

\bibitem{burgess} 
C.P. Burgess, J. Cline, E. Filotas, J. Matias, G.D. Moore,
``Loop-Generated Bounds on Changes to the Graviton Dispersion
Relation'', {\bf JHEP} 0203 (2002) 043

\bibitem{Chaichian:2000si}
M.~Chaichian, M.~M.~Sheikh-Jabbari and A.~Tureanu,
``Hydrogen atom spectrum and the Lamb shift in noncommutative QED,''
{\it Phys.\ Rev.\ Lett.}  {\bf 86}, 2716 (2001)
[arXiv:hep-th/0010175].
\\
Z.~Guralnik, R.~Jackiw, S.~Y.~Pi and A.~P.~Polychronakos,
``Testing non-commutative QED, constructing non-commutative MHD,''
{\it Phys.\ Lett.} B {\bf 517}, 450 (2001)
[arXiv:hep-th/0106044].

\bibitem{NCOS-charged}
C.~S.~Chu,
``Noncommutative open string: Neutral and charged,''
arXiv:hep-th/0001144.

\bibitem{NC-QL}
A.~Anisimov, T.~Banks, M.~Dine and M.~Graesser,
``Comments on non-commutative phenomenology,''
{\it Phys.\ Rev.} D {\bf 65}, 085032 (2002)
[arXiv:hep-ph/0106356].
\\
G.~Amelino-Camelia, G.~Mandanici and K.~Yoshida,
``On the IR / UV mixing and experimental limits on the parameters of 
canonical noncommutative spacetimes,''
arXiv:hep-th/0209254.
\\
C.~E.~Carlson, C.~D.~Carone and R.~F.~Lebed,
``Supersymmetric noncommutative QED and Lorentz violation,''
arXiv:hep-ph/0209077.
\\
A.~Armoni and E.~Lopez,
``UV/IR mixing via closed strings and tachyonic instabilities,''
{\it Nucl.\ Phys.} B {\bf 632}, 240 (2002)
[arXiv:hep-th/0110113].

\bibitem{NC-LE}
I.~Mocioiu, M.~Pospelov and R.~Roiban,
``Low-energy limits on the antisymmetric tensor field background on the 
brane and on the non-commutative scale,''
{\it Phys.\ Lett.} B {\bf 489}, 390 (2000)
[arXiv:hep-ph/0005191].
\\
I.~Mocioiu, M.~Pospelov and R.~Roiban,
``Limits on the non-commutativity scale,''
arXiv:hep-ph/0110011.

\bibitem{denisov}

Denisov, Victor, ``New effect in nonlinear Born-Infeld
electrodynamics'', {\it Phys. Rev. D.} {\bf 61}, 036004, 2000.

\bibitem{magnet}

{\it 2001 National High Magnetic Field Laboratory (NHMFL) Annual
Research Review}. NHMFL Publications. (www.magnet.fsu.edu) (2001).

\bibitem{lantz}

Lantz, B. {\it et al.}, ``Quantum limited optical phase detection at
the $10^{-10}$-rad level'' {\it Optical Society of America.} Vol. 19,
No. 1, pp 91. 2002.

\bibitem{ligo}

Barish, B. and Weiss, R., ``LIGO and the detection of gravitational
waves.'', {\it Phys. Today}, 52, 44-50 (1999)

\bibitem{enqvist} Enqvist, Kari. ``Primordial Magnetic Fields'', {\it
Int. J. Mod. Phys.} D7 (1998) 331-350.

\bibitem{riis} Riis, E. {\it et al.,} ``Test of the Isotropy of Light
Using Fast-Beam Laser Spectroscopy'', {\it Phys. Rev. Lett.} 60. 81-84
(1988)

\end{thebibliography}
\end{document}